# Frequency Response Analysis of Backside Illuminated Free-Membrane Type IR Superconducting Bolometer

R. Mohajeri, S. I. Mirzaei, and M. Fardmanesh, *Member IEEE*

*Abstract*— The IR response of a backside illuminated type Yttrium Barium Copper Oxide (YBCO) transition-edge bolometer (TEB) in a free-membrane configuration is measured and analyzed. The analysis is based on a comprehensive equivalent circuit model for the bolometer structure considering all the thermal parameters of the device. The detector is made using a 450nm superconductive YBCO thin film grown on about 100μm thick crystalline SrTiO$_3$ (STO) substrate using an optimized low cost Metal Organic Deposition (MOD) technique. The backside illuminated structure provided a basis for freely utilizing different IR absorber materials and structures on the backside of the bolometer in order to get wide band high radiation absorption coefficient for the device, while also avoiding any negative effect of the deposition of the absorber layer on the YBCO thin film properties. The obtained analytical and experimental results for the backside illuminated bolometer were also compared with the results of the previously reported front-side illuminated bolometers.

## I. Introduction

THE voltage response of superconducting transition edge bolometers (TEB) versus radiation modulation frequency has previously been investigated in different studies where the superconducting thin film was exposed directly to the incoming radiation, so that the radiation energy was mainly absorbed by the superconducting thin film [1-3]. There have also been some efforts to study the effect of different thermal parameters on the responsivity of this type of detectors, by using different thermal models [3-6]. These models have provided the optimization of the thermal parameters of the TEB to get high responsivity from the detector [7-13].

In this paper, we propose a backside illuminated free-membrane type superconducting transition edge bolometer, which provides a basis for freely utilizing different absorber materials and structures on the backside of the bolometer in order to get wide band high radiation absorption coefficient for the device. This is while this configuration allows avoiding any negative effect of the deposition of the absorber layer on the YBCO thin film properties. The obtained experimental results for the backside illuminated bolometer were compared with the results of the previously reported front-side illuminated bolometers.

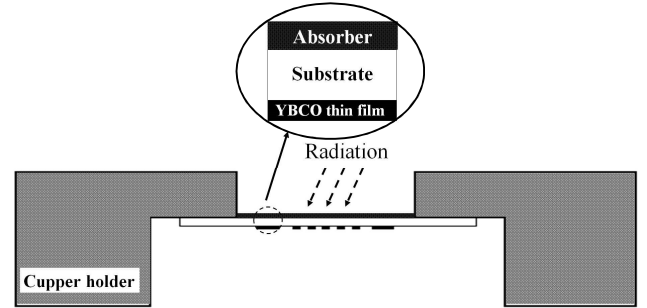

Fig. 1. The schematic of free-membrane bolometer structure, illuminated from the backside

A one-dimensional comprehensive thermal model is also proposed for this configuration to analytically explain the device performance versus the modulation frequency, which is determined by:

$$f = \frac{D}{\pi L^2}, \qquad (1)$$

where, L is the thermal diffusion length into the substrate, D = $K_s/C_s$ is the thermal diffusivity of the substrate, and $K_s$ and $C_s$ are the thermal conductivity and the specific heat capacity(per unit volume) of the substrate material, respectively.

## II. Device preparation

The sample studied in this paper was made of a 450nm thick superconducting YBCO thin film, deposited on a 100μm single crystalline STO substrate using cost Metal Organic Deposition (MOD) technique (see [14], [15] for details). Then, the deposited YBCO thin film was patterned using standard photolithography technique (positive photoresist) with 55μm wide and 1.6mm long meander line pattern providing a total area of 1.9mm$^2$ for the device structure. In order to increase the IR absorption of the bolometer, a very thin absorber layer of black marker was applied on the backside of the substrate. The fabricated detector was mounted on a copper sample holder with

R. Mohajeri (e-mail: roya.mohajeri@gmail.com), and M. Fardmanesh (e-mail: fardmanesh@sharif.edu) are with Superconducting Electronics Research Laboratory, Sharif University of Technology, Tehran. Iran.
S.I. Mirzaei (e-mail: imanmirzaie@gmail.com) is with the Faculty of Basic Sciences, Tarbiat Modares University, Tehran, Iran.

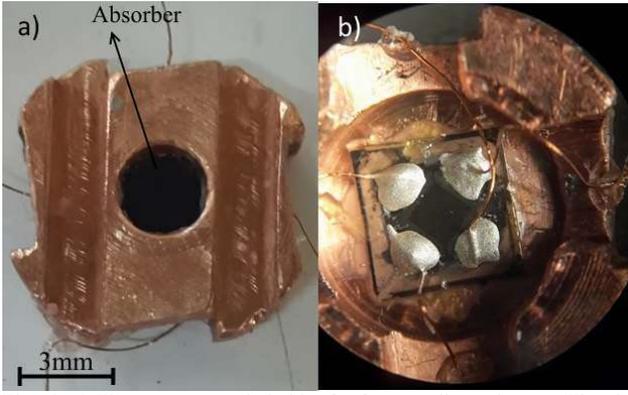

Fig. 2. a) The copper sample holder for free-standing substrate (The absorber layer on the backside of the device is seen at the middle of the sample holder), b) the fabricated bolometer mounted on the sample holder (electrical contact was implemented using silver paste and copper wires, and the absorber was applied on the backside of the device).

centered mounting hole, providing a free-standing configuration for the bolometer.

The schematic structural description of the device is illustrated in Fig.1. According to this structure, the device is exposed to the radiation power from the backside, where the absorber layer is applied. By using this configuration, all possible negative effects of the absorber material on the superconducting thin film quality are avoided, while it is possible to utilize this kind of detectors for all the radiation frequencies including terahertz (THz) and infrared (IR) ranges.

Fig. 2 shows the fabricated bolometer, mounted on a copper holder. The copper holder structure provides a free standing configuration for the thin device substrate in order to increase the responsivity of the detector due to having lower thermal conductivity to the copper holder [13]. The electrical contacts to the bolometer were made by copper wires and silver epoxy treated on $80^0C$ hot plate on the gold coated contact areas of the pattern. The device was exposed to a 573K blackbody source from the absorber side and the radiation of the source is chopped using a mechanical chopper unit. The output voltage of the detector was then measured using a lock-in amplifier based low noise characterization setup (see [16] for details).

### III. ANALYTICAL THERMAL MODEL

Considering a large area superconducting pattern compared to the substrate thickness, a one-dimensional analytical thermal model is proposed with all the thermal parameters defined per unit area. In this model, the thermal boundary conductance of the substrate/copper-holder is modeled with a constant electrical conductance. This is while, the thermal properties of the substrate, considered as a continuous medium, are modeled with the equivalent distributed electrical resistance and capacitance. As the thickness of the superconducting film in our sample was very low compared to the substrate thickness, a lumped-element assumption for the superconducting thin film was allowed for the considered modulation frequency range in the model.

The analytical thermal model of the fabricated device is shown in Fig. 3, where Q is the absorbed radiation intensity at the absorber surface. In this model, the thermal conductivity of both the superconducting thin film and the absorber layer were found to be negligible in the measured modulation frequencies, due to their very low thickness compared to that of the substrate. Also, the thermal conductivity at the boundaries of the absorber/substrate and the superconducting thin film/substrate was very high compared to the thermal parameters of the 100µm thick substrate material, which resulted in a negligible effect on the voltage response of the device as confirmed in both the simulation and the measurement results.

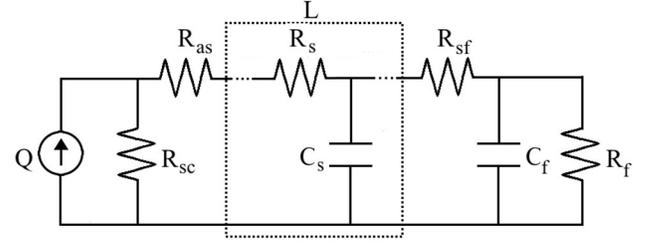

Fig. 3. Equivalent thermo-physical circuit model of the sample in contact with the copper holder, where Q is the input radiation power, $C_f$ is the lumped heat capacitance of the film, Cs and Rs are the heat capacity and the thermal resistivity of the substrate material, respectively, $R_{sf}$, $R_{as}$ and $R_{sc}$ are the thermal boundary resistance values at the substrate/YBCO film, absorber/substrate and the substrate/coper holder interfaces, respectively.

Considering the thermal model of Fig. 3, at lower modulation frequencies, the temperature variation at the superconducting thin film is considerably affected by the absorbed radiation power at the backside of the detector. This is explained by the fact that the thermal diffusion length into the substrate is more than the substrate thickness at low modulation frequencies. There is a frequency limit, called the "knee frequency" [1], [16], at which the thermal diffusion length into the device equals to the substrate thickness. The knee frequency could be obtained using equation (1) by setting the thermal diffusion length with the substrate thickness. At the modulation frequencies above the knee frequency, a drop in the slope of the voltage response is expected, which is confirmed by the obtained measurement and analytical results as discussed in the following section.

### IV. RESULTS AND DISCUSSION

The absorption coefficient of the 450nm thick YBCO thin film on STO substrate was measured in the wavelength range between 5µm and 25µm, which is shown in Fig. 4. According to the results, the fabricated YBCO film is highly reflective at the mid-IR frequencies, which necessitates using a proper absorber layer on the detector. Carbon black and black marker have shown an absorption coefficient above 0.7 at the mid-IR frequencies (Fig. 4), which is three times higher than YBCO film. Therefore, black marker was implemented on fabricated device, as an absorber layer to increase its sensitivity to the incident radiation power. To avoid the negative effects of the black marker on the superconducting properties of the YBCO film, the absorber layer was implemented on the backside of the device, right on the STO substrate. The absorption coefficient measurements were performed by a Bruker Vertex 70 FTIR

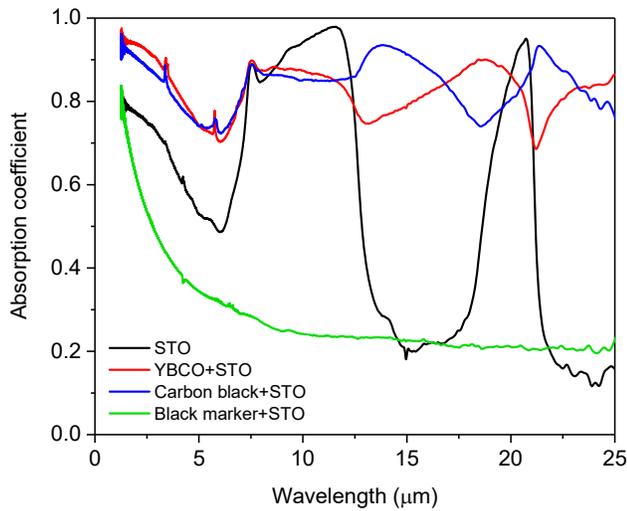

Fig. 4. The absorption coefficient of STO substrate, and superconducting YBCO thin film, carbon black, and black marker on STO substrate.

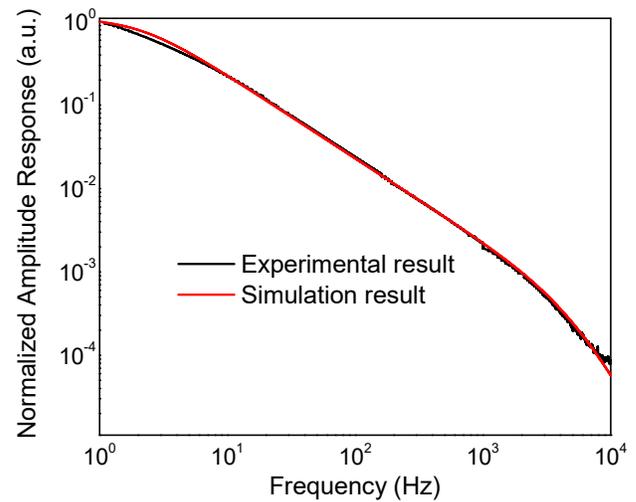

Fig. 6. The simulated and measured voltage amplitude response of the fabricated free-membrane bolometer.

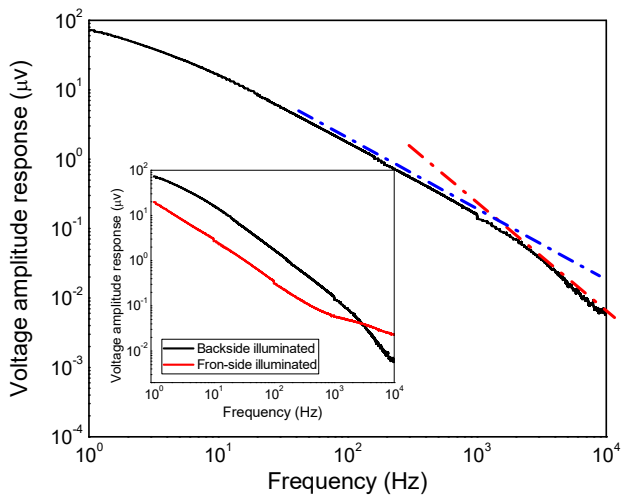

Fig. 5. The voltage amplitude response of the fabricated free-membrane bolometer. Inset: the voltage response of the fabricated bolometer in black: backside illuminated, and red: front-side illuminated case.

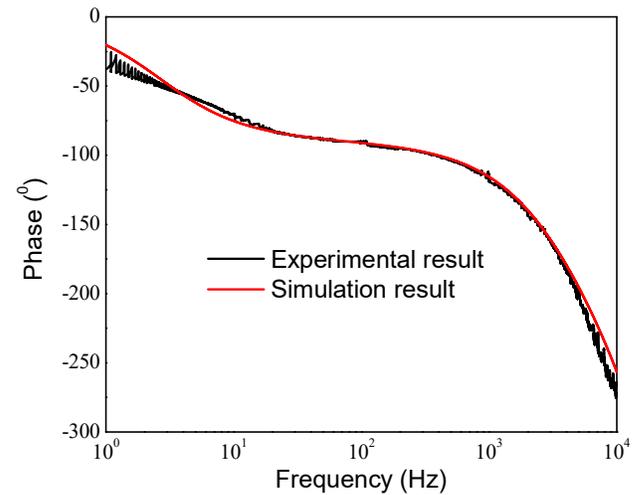

Fig. 7. The simulated and measured voltage phase response of the fabricated free-membrane bolometer.

spectrometer.

The voltage amplitude response versus the modulation frequency of the free-membrane bolometer, illuminated from the backside by a 573K blackbody source is shown in Fig. 5. According to the result, the voltage response drops steeply above 2 KHz modulation frequency, where the thermal diffusion length is equal to the substrate thickness, and therefore the effect of the absorber layer on the voltage response starts to get eliminated. The inset of Fig. 5 shows the voltage response of the detector with backside illuminated and front side illuminated configuration. As it is shown in the inset of Fig. 5, the backside illuminated structure is highly responsive at the modulation frequencies below 5 KHz compared to the front side illuminated structure, which is due to the use of the black marker, simply applied on the STO substrate. This is while, at the front side illuminated configuration, the YBCO film is influenced by the absorbed radiation power at the film surface for all modulation frequencies. The backside illumination of the device, however, provides a basis for the detection of a wide range of wavelengths, in which the superconducting YBCO thin film is highly reflective.

Using the proposed one dimensional thermal model in section III, the responsivity of the backside illuminated free-membrane device was also analysed. Fig. 6 shows the simulated and measurement result of the voltage amplitude response of the fabricated device. According to the results in Fig. 6, only a small discrepancy is evident between the experimental and simulated curves which indicates that the one dimensional thermal model well investigates the heat flow mechanism caused by the absorbed radiation at the device surface. The voltage phase response obtained from the model was also found to be in agreement with the measurement observations, as is shown in Fig. 7.

According to the model, the bolometer would dissipate the absorbed radiation power laterally through the thermal contact between the substrate/copper-holder boundary resistance, and also across the substrate thickness. At the very low modulation frequencies, the heat distribution is mostly through the substrate/copper holder boundary, resulting in a decreasing phase response as a function of the modulation frequency. Above 200Hz modulation frequency, the absorbed radiation power dissipation has considerable distribution through the substrate thickness, resulting in a plateau in the phase of the response [4]. The thermal parameters of superconducting film result in a decrease in the phase of the response at the higher modulation frequencies above the knee frequency.

## V. Conclusion and summary

A free-membrane type superconducting bolometer with backside illumination configuration was fabricated and its responsivity to a mechanically chopped blackbody source was investigated. The backside illumination configuration for this device allowed the application of desired absorber layers for wide band functionality of the detector without destroying the properties of the superconducting YBCO film. To increase the sensitivity of the device at IR wavelengths ranging from 5µm to 25 µm, a black marker was utilized as the absorber layer at the backside of the detector. An equivalent thermophysical model was also proposed for the device structure to analytically explain the device performance versus the modulation frequency. The results obtained from the model has shown a small discrepancy with that of the measurement which indicates that the proposed model well investigates the heat flow mechanism in the device structure.